\documentstyle{lamuphys}
\makeatletter
\let\chapter\hid@chapter
\makeatother
\begin{document}
\pagenumbering{arabic}
\title{AAO support observations for the Hubble Deep Field South}

\author{B.J.Boyle}

\institute{Anglo Australian Observatory, PO Box 296, Epping, NSW 2121}

\maketitle

\begin{abstract}
We present proposed ground-based support observations at the AAO for
the forthcoming Hubble Deep Field South (HDF-S) campaign.

\end{abstract}
\section{Introduction}
In October 1998, the Hubble Space Telescope (HDF-S) will once again
spend several hundred hours imaging an area of sky to yield a deep
image of the distant Universe.  The region was chosen to lie in the
Southern Continuous Viewing Zone (CVZ) at declination $-61^{\circ}$
and includes a $z=2.24$ QSO (Q2233.6-6033), first identified by the UK
Schmidt Telescope and confirmed spectroscopically by the 2dF
instrument at the Anglo-Australian Telescope (Boyle 1997, see also
Sealey et al.\, these proceedings).  In the HST campaign, the STIS
observations will be centred on the QSO, with the WFPC2 deep field
offset by 9 arcmin west.  Positions of the HST fields are given in
table 1.  Full details of the HST observations may be obtained from
the STScI website {\tt http://www.stsci.edu/}.

\begin{table}[htb]
\caption[ ]{HDF-S positions}
\begin{center}
\renewcommand{\arraystretch}{1.2}
\begin{tabular}{lc}
\hline\noalign{\smallskip} 
Instrument&RA (2000) Dec\\
{\bf WFPC2  }&22 32 56.2 --60 33 03\\
{\bf STIS  }&22 33 37.7 --60 33 29\\
{\bf NICMOS3 \qquad }&22 32 52.4 -60 38 33\\
\noalign{\smallskip}
\hline
\end{tabular}
\renewcommand{\arraystretch}{1}
\end{center}
\end{table}

\section{AAO observations}

The HST observations provide a unique opportunity for Southern
Hemisphere facilities to provide essential ground-based support for
the HDF-S.  Although there is currently no operational 8m-class
telescope in the South, the existing 4m-class telescopes can still
play a major role in this early stage of the HDF-S campaign.  We
detail below the proposed observations which the Anglo-Australian
Telescope will under in support of the HDF-S. All data obtained will
be made available to the community as quickly as possible, hopefully
timed to co-incide with the release of the HST imaging and
spectroscopic data.  A WWW page has been set-up at the AAO {\tt
http://www.aao.guv.au/hdfs/} to distribute information and data 
relating to
these proposed observations.

\begin{enumerate}

\item {\bf Echelle observations of Q2233.6-6033}.  UCLES observations
of the QSO ($B=17.5$) at the centre of the STIS field will be used to
generate a list of absorption lines with $W_{\lambda}> 24$m\AA\
(3$\sigma$) over the wavelength region 3334\AA--5045\AA\, corresponding
to Ly$\beta$ -- CIV at the redshift of the QSO.  Observations are 
planned for July.

\item {\bf Intermediate-depth prime-focus imaging}.  Broadband $BRI$
imaging will be used to generate an catalogue of $R<24$ objects in a
$9 \times 9\,$arcmin region comprising both the STIS and WFPC fields.
Observations will be obtianed in May and used to provide an input
catalogue for the LDSS++ observations planned in July.

\item {\bf Spectroscopy of $R<24$ galaxies in the HDF-S}.  An
innovative upgrade to the LDSS instrument at the AAT (LDSS++,
Glazebrook et al.\ 1998) will enable spectra for up to 300 faint
($R<24$) galaxies to be obtained over a $9 \times 3\,$arcmin region
comprising both the WFPC and STIS deep fields.  LDSS++ observations
are currently planned for July.

\item {\bf Flanking field QSO absorption line systems}.  2dF
observations of colour and/or prism-selected QSO candidates in the
3-deg$^2$ region centred on the HDF-S have been proposed by
Hewett {\it et al.}  These observations would be used to derive
information on the structure of QSO absorption line systems on the
largest possible scales.

\end{enumerate}

In addition, the Australia Telescope Compact Array (PI: Norris) 
will be used to construct a deep radio map (few $\mu$Jy rms) 
of the HDF-S at 13cm.  Initial ATCA observations (at 6 and 20cm) 
of the field have already been obtained confirming the lack of any
strong ($>100$mJy) radio sources near the HDF-S.

%

%
%

\end{document}